# Emergent excitations in a geometrically frustrated magnet

S.-H. Lee[*], C. Broholm†[*], W. Ratcliff‡, G. Gasparovic†, Q. Huang[*], T.H. Kim‡∥, & S.-W. Cheong‡

[*] *NIST Center for Neutron Research, National Institute of Standards and Technology, NIST, Gaithersburg, Maryland 20899, USA*

† *Department of Physics and Astronomy, The Johns Hopkins University, Baltimore, Maryland 21218, USA*

‡ *Department of Physics and Astronomy, Rutgers University, Piscataway, New Jersey 08854, USA*

∥ *Present address: Francis Bitter Magnet Lab, Massachusetts Institute of Technology, Cambridge, Massachusett 02139, USA*

**Frustrated systems are ubiquitous[1,2,3] and interesting because their behavior is difficult to predict. Magnetism offers extreme examples in the form of spin lattices where all interactions between spins cannot be simultaneously satisfied.[4] Such "geometrical frustration" leads to macroscopic degeneracies, and offers the possibility of qualitatively new states of matter whose nature has yet to be fully understood. Here we have discovered how novel composite spin degrees of freedom can emerge from frustrated interactions in the cubic spinel $ZnCr_2O_4$. Upon cooling, groups of six spins self-organize into weakly interacting antiferromagnetic loops whose directors, defined as the unique direction along which the spins are aligned parallel or antiparallel, govern all low temperature dynamics. The experimental evidence comes from a measurement of the magnetic form factor by inelastic neutron scattering. While the data bears no resemblance to the atomic form factor for chromium, they are perfectly consistent with the form factor for hexagonal spin loop directors. The hexagon directors are to a first approximation**

**decoupled from each other and hence their reorientations embody the long-sought local zero energy modes for the pyrochlore lattice.**

Magnetism in transition metal oxides stems from atomic spins on the vertices of a periodic lattice. In insulators, interactions generally favor anti-parallel nearest neighbor spin alignment. For a simple cubic lattice (Fig. 1(a)), only a long-range ordered spin configuration can satisfy all interactions. On cooling, such systems show a continuous increase in the spin correlation length culminating in a phase transition to long-range order. However, for spins on the vertices of corner-sharing tetrahedra (Fig.1(b)), no configuration can satisfy all interactions; a magnetic predicament labeled "geometrical frustration.[4]" Because the spin interaction energy is minimized when the four spins on each tetrahedron add to zero, interactions do not call for a divergent correlation length, but simply define a restricted phase space for fluctuations, parametrized by $\theta$ and $\phi$ for each tetrahedron. [5,6] Just as composite fermions can emerge from degenerate Landau levels in a two-dimensional electron gas[7], the near degenerate manifold of states in a frustrated magnet is fertile ground for emergent behavior[8].

While direct structural probes are unavailable for quantum hall samples, we can monitor the correlated spin state in a frustrated magnet by scattering neutrons from it.[9] Neutrons carry a magnetic dipole moment and are subject to forces from atomic scale field-gradients. The resulting pattern of quasi-elastic scattering versus wave vector transfer, $\mathbf{Q}=\mathbf{k}_i-\mathbf{k}_f$, is the sample averaged Fourier transform of spin configurations within a coherence volume of order (100 Å)$^3$ given by the instrumental resolution. Here $\mathbf{k}_i$ and $\mathbf{k}_f$ are the de-Broglie wave vectors associated with the incident and scattered neutrons respectively. Magnetic peaks generally sharpen with decreasing temperature as the correlation length, $\xi$, increases. For un-frustrated $La_2CuO_4$[10], the half width at half maximum (HWHM), $\kappa(T) = \xi(T)^{-1}$, becomes indistinguishable from zero below a microscopic energy scale, the Curie-Weiss temperature $|\Theta_{CW}|$. In contrast, for frustrated $ZnCr_2O_4$, as shown in Fig. 2, $\kappa$ remains finite even below $|\Theta_{CW}|$ and extrapolates to a finite value as $T/|\Theta_{CW}| \to 0$.

The finite low temperature correlation length in $ZnCr_2O_4$ signals the emergence of confined nano-scale spin clusters. Rather than being associated with temperature dependent short range order above a phase transition, the wave-vector dependence of the low $T$ intensity shown in Fig. 3(a) and Fig. 3(b) can be interpreted as a spin cluster form factor. As opposed to the form factor for an atomic spin, the cluster form-factor vanishes for $Q \to 0$, which is evidence that clusters carry no net spin[11]. Further analysis is complicated by the anisotropy of spin clusters, which can occur in four different orientations for a cubic crystal. Rather than Fourier-inverting the data, we therefore compare them to the orientationally averaged Fourier transform of potential spin clusters. Individual tetrahedra would be prime candidates as they constitute the basic motif of the pyrochlore lattice. However, a tetrahedron is too small to account for the features observed in Fig. 3(a)-(b). The next smallest symmetric structural unit is the hexagonal loop formed by a cluster of six tetrahedra [Fig. 4 (a)]. Two spins from each tetrahedron occupy the vertices of a hexagon while the other two spins from each tetrahedron belong to different hexagons. Averaging over the four different hexagon orientations in the pyrochlore lattice, the antiferromagnetic hexagon spin loop form factor squared can be written as[22]

$$|F_6(\mathbf{Q})|^2 \propto \{\sin\frac{\pi}{2}h \cdot (\cos\frac{\pi}{2}k - \cos\frac{\pi}{2}l)\}^2$$
$$+ \{\sin\frac{\pi}{2}k \cdot (\cos\frac{\pi}{2}l - \cos\frac{\pi}{2}h)\}^2$$
$$+ \{\sin\frac{\pi}{2}l \cdot (\cos\frac{\pi}{2}h - \cos\frac{\pi}{2}k)\}^2$$

The magnetic neutron scattering intensity would follow $I_0(\mathbf{Q}) = |F_6(\mathbf{Q})|^2 |F(Q)|^2$ where $F(Q)$ is the magnetic form factor for $Cr^{3+}$. The excellent qualitative agreement between model and data in Fig. 3 provides compelling evidence that neutrons scatter from antiferromagnetic hexagonal spin clusters rather than individual spins. In effect,

ZnCr$_2$O$_4$ at low temperatures is not a system of strongly interacting spins, but a protectorate[12] of weakly interacting spin-loop directors. Thermal and quantum fluctuations that violate collinearity within the hexagons should induce residual interactions between directors. Such interactions may account for the inelasticity of the scattering, the director correlations reflected in the greater sharpness of the experimental features in Figs. 2 and 3, and the increase of $\kappa$ with $T$, which indicates gradual disintegration of the directors.

What is the basis for the emergence of spin loop directors as the effective degrees of freedom in this frustrated magnet? Fig. 4 shows the spins surrounding a hexagon in the pyrochlore lattice. Spin configurations that satisfy all interactions are characterized by the connected vectors of Fig. 4 (b). While the spin configuration remains severely under-constrained, no adjustment is possible without affecting spins on the outer perimeter. A general lowest energy spin configuration on the pyrochlore lattice should therefore have wave-like soft modes that extend throughout the lattice. However, if the six hexagon spins are anti-parallel with each other as shown in Fig. 4 (c), then the staggered magnetization vector for a single hexagon, which for brevity shall be denoted the spin loop director, is decoupled from the 12 outer spins and hence its reorientation embodies a local zero energy mode for the pyrochlore lattice[13]. Remarkably, it is possible to assign all spins on the spinel lattice to hexagons simultaneously thus producing $N/6$ weakly interacting degrees of freedom (see Fig. 1 (c)). Accordingly, states that account for 1/6 of the entropy of the magnet are accessible through local fluctuations from configurations where all spins are bunched into directors. Indeed, the measured entropy of 0.15 $R$ln4 just above a low temperature spin-Peierls like phase transition[4,14] is close to the predicted entropy of (1/6) $R$ln4 for uncorrelated directors in a spin-3/2 magnet. In contrast, there are *no* local soft modes for a general spin configuration in the low energy manifold. This distinction between a general state from the low energy manifold and states from the director protectorate is likely the basis for the stability of the latter.

Our finding that fluctuations in $ZnCr_2O_4$ involve spin clusters and not individual spins, provides a natural explanation for a range of surprising properties of geometrically frustrated magnets. It explains why the temperature dependent susceptibility of frustrated magnets is accurately described by exact diagonalization of judiciously chosen spin clusters[15-17]. Moreover, the so-called "undecouplable" $\mu$SR response[18] is recognized as being a consequence of muons sensing slow large amplitude fluctuations of select spins in a spin director. A director protectorate also provides a natural explanation for the coexistence of low characteristic energy scales with rigid short-range order as evidenced by specific heat [19] and quasi-elastic neutron data[20].

The director protectorate hints at an organizing principle for frustrated systems: If macroscopic condensation is not possible, interacting degrees of freedom combine to form rigid composite entities or clusters with weak mutual interactions. Exploring the generality and basis for such a principle should be an interesting focus for theoretical work and experiments on a wider class of systems. Composite degrees of freedom are common in strongly interacting many body systems. Quarks form hadrons, hadrons form nuclei, nuclei plus electrons form atoms, atoms form molecules, which in turn are the basis for complex biological functionality. Planets, stars, galaxies and galaxy clusters are examples of clustering on a grander length scale. However, to our knowledge, the emergence of a confined spin cluster degree of freedom has not previously been documented in a uniform gapless magnet. The discovery is important because magnets offer an opportunity not afforded by the aforementioned systems to monitor emergent structure in complex interacting systems with microscopic probes such as neutron scattering and NMR. The collapse of a geometrically frustrated magnet into a director protectorate could for example be a useful template for exploring aspects of protein folding[2,21].

**Methods**

Three crystals of $ZnCr_2O_4$ with a total mass of 200 mg, were co-mounted for the inelastic neutron scattering measurements. The measurements were performed using the cold neutron triple-axis spectrometer SPINS at the National Institute of Standards and Technology Center for Neutron Research. A vertically focusing pyrolytic graphite (002) monochromator [PG(002)] extracted a monochromatic beam with energy $E_i$ = 6.1 meV from a $^{58}$Ni coated cold neutron guide. Scattered neutrons were analyzed with seven or eleven 2.1 cm × 15 cm PG(002) analyzer blades that reflected neutrons with $E_f$ = 5.1 meV onto a $^3$He proportional counter. Cooled Be filtered the scattered beam.

There it accounts for the structure factor of a nonzero-energy local antiferromagnon on ferromagnetically aligned hexagonal spin loops.

**Acknowledgements**

We thank O. Tchernyshyov, R. Moessner, S. L. Sondhi, A. B. Harris, G. Aeppli, N. Read, and D. Weitz for helpful discussions, J. J. Rush, A. P. Ramirez and P. M. Gehring for critical reading of the manuscript, and Z. Huang for assistance in making figures. This work was partially supported by the NSF and the BSF.

Correspondence and requests for materials should be addressed to S.H.L. (e-mail: shl@nist.gov).

**Figure captions**

**Figure. 1.** Lowest energy spin configurations for four antiferromagnetically interacting spins on a square and a tetrahedron, and the pyrochlore lattice of corner-sharing tetrahedra. **a**, Setting aside global rotations, four spins on the vertices of a square with nearest neighbor interactions have a unique lowest energy spin configuration. **b**, For four spins on the vertices of a tetrahedron, any configuration with vanishing total spin has the lowest configuration energy. **c**, The lattice of corner-sharing tetrahedra formed by the octahedrally coordinated B sites in a spinel structure with chemical formula $AB_2O_4$. A periodic assignment of all spins in the pyrochlore lattice is made to four different types of non-overlapping hexagons, represented by the colors blue, green, red, and gold. Every spin belongs to just one hexagon and each such hexagon carries a six spin director. The resulting tetragonal structure of these hexagons has a unit cell of $2a \times 2a \times 3c$ and can be described by a stacking of two different types of three-layer slabs along the c-axis. The hexagon coverage on consecutive slabs is in fact uncorrelated, so that a macroscopic number of random slab-sequences can be generated.

**Figure 2.** Temperature dependence of the inverse correlation length, $\kappa(T) = \xi(T)^{-1}$. The data were derived from antiferromagnetic neutron scattering

peaks by fitting to resolution-convoluted lorentzians. The triangles and circles are the lorentzian HWHMs along the $\left(h\frac{5}{4}\frac{5}{4}\right)$ direction for $\hbar\omega$ = 1 meV, obtained with seven and eleven analyzer blades, respectively. $\kappa$ does not vanish as $T/|\Theta_{CW}| \to 0$, but extrapolates to a value that is close to the HWHM associated with the squared form factor for antiferromagnetic hexagon spin loops (dashed line). The inset shows raw T=15 K data for $ZnCr_2O_4$. The bar shows the experimental resolution. The solid line is a resolution convoluted 2D-lorentzian, the dashed line is the squared hexagon spin loop form factor convoluted with resolution.

**Figure 3.** Wave vector dependence of the inelastic neutron scattering cross section for $ZnCr_2O_4$. **a-b**, Color images of inelastic neutron scattering intensities from single crystals of $ZnCr_2O_4$ in the (hk0) and (hkk) symmetry planes obtained at T=15K for $\hbar\omega$ = 1 meV. The data are a measure of the dynamic form factor for self-organized nano-scale spin clusters in the material. **c-d**, Color images of the form factor squared calculated for antiferromagnetic hexagon spin loops averaged over the four hexagon orientations in the spinel lattice. The excellent agreement between model and data identifies the spin clusters as hexagonal spin loops.

**Figure 4.** Possible spin fluctuations in the classical ground state manifold. **a**, Spin cluster surrounding a hexagon (shown in yellow) in the pyrochlore lattice of Fig. 1**c**. **b**, Generalized pattern of classical spin vectors on six neighboring tetrahedra satisfying the condition that there be no net spin on any tetrahedron. **c**, Pattern of spin vectors satisfying the condition that $\theta$=0 for all tetrahedra. For such spin configurations, each tetrahedron has two pairs of antiparallel spins, and each hexagon has six collinear and antiferromagnetic spins as indicated by six red arrows in panel **a**. The energy of the spin cluster is independent of the orientation of the spin loop directors.